\documentclass[10pt,twocolumn]{IEEEtran}

\usepackage[cmex10]{amsmath}
\usepackage{mathtools}
\usepackage{amssymb}
\usepackage{array}
\usepackage[pdftex]{graphicx}
\usepackage{multirow}
\usepackage{cases}
\usepackage[tight,footnotesize]{subfigure}
\usepackage{algorithm} %format of the algorithm
\usepackage{algorithmic} %format of the algorithm
\usepackage{booktabs}
\usepackage{graphicx}
\usepackage{url}
\newtheorem{theorem}{\textbf{Theorem}}

\newcolumntype{C}[1]{>{\vspace{0.5em}\begin{minipage}{#1}\centering\let\newline\\
\arraybackslash\hspace{0pt}}m{#1}<{\end{minipage}\vspace{0.5em}}}

\usepackage{nomencl}
\usepackage{cite}
\usepackage{xspace}
\newcommand{\ouralg}{UPMR\xspace}

\makenomenclature
\begin{document}
%\title{Pricing Tenant Scheduling Flexibilities to Reduce
%Energy Costs in Cloud Data Centers}

\title{\fontsize{23pt}{26pt}\selectfont{Extending Demand Response to Tenants in Cloud Data Centers via Non-intrusive Workload Flexibility Pricing}}

\author{Yong~Zhan,~\IEEEmembership{Student Member,~IEEE,}
        Mahdi~Ghamkhari,~\IEEEmembership{Student Member,~IEEE,}
        Du~Xu,~\IEEEmembership{Member,~IEEE,}
        Shaolei~Ren,~\IEEEmembership{Member,~IEEE,}
        and~Hamed~Mohsenian-Rad,~\IEEEmembership{Senior Member,~IEEE} \vspace{-0.3cm}

 \thanks{Y. Zhan and D. Xu are with the Key Laboratory of Optical Fiber Sensing and Communications, University of Electronic Science and Technology of China, Chengdu, China. M. Ghamkhari, S. Ren, and H. Mohsenian-Rad are with the Department of Electrical Engineering, University of California, Riverside, CA, USA. This work was done when Y. Zhan was a Visiting Student at the Smart Grid Research Lab, University of California at Riverside. %The % corresponding author is H. Mohsenian-Rad.
 } \vspace{-0.4cm}}

\maketitle

\begin{abstract}
%\boldmath
Participating in demand response programs is a promising tool for reducing energy costs in data centers by modulating
energy consumption. Towards this end, data centers can employ
a rich set of resource management knobs, such as
workload shifting and dynamic server provisioning.
Nonetheless, these knobs may \emph{not}
be readily available in a cloud data center (CDC)
that serves cloud tenants/users, because workloads in CDCs are
managed by tenants themselves who are typically charged based on a usage-based or flat-rate pricing and
often have no incentive to cooperate with the CDC operator for demand response and cost
saving. Towards breaking such ``split incentive'' hurdle, a few recent studies have tried market-based mechanisms, such as dynamic pricing, inside CDCs. However, such mechanisms often rely on complex designs that are hard to implement and difficult to cope with by tenants. To address this limitation, we propose a novel incentive mechanism that is not dynamic, i.e., it keeps pricing for cloud resources unchanged for a long period. While it charges tenants based on a Usage-based Pricing (UP) as used by today's major cloud operators, it rewards tenants proportionally based on the time length that tenants set as deadlines for completing their workloads. This new mechanism is called Usage-based Pricing with Monetary Reward (\ouralg).
%
%which provides monetary incentives to
%encourage cloud tenants' workload shifting/delaying
%over time and allows the CDC operator to flexibly
%process these workloads in a cost-efficient manner.
%
We demonstrate the effectiveness of \ouralg
both analytically and empirically. We show that \ouralg can reduce
the CDC operator's energy cost
by 12.9\% while increasing its profit by 4.9\%,
compared to the state-of-the-art approaches used by today's CDC
operators to charge their tenants.

\end{abstract}


\IEEEpeerreviewmaketitle

\nomenclature{$i$}{Index for type of users.}
\nomenclature{$j$}{Index for type of demand charge.}
\nomenclature{$t$}{Index for time slot.}
\nomenclature{$k$}{Index for type of instance.}
\nomenclature{$I$}{Number of types of users.}
\nomenclature{$K$}{Number of types of instance.}
\nomenclature{$J$}{Number of types of demand charge.}
\nomenclature{$A_j$}{Set of time slots related to type $j$ demand charge.}
\nomenclature{$T$}{Length of each time slot.}
\nomenclature{$\tau$}{Number of time slots.}
\nomenclature{$\Psi$}{Unit number of requests.}
\nomenclature{$\kappa_i$}{Loss factor of $i$.}
\nomenclature{$D_{max}$}{Maximum threshold to time-shift workloads.}
\nomenclature{$\delta$}{Price of CDC resources for processing $\Psi$ requests.}
%\nomenclature{$\theta_{i,k}[t]$}{Number of type $k$ instances rented by $i$ at $t$.}
\nomenclature{$\lambda_i[t]$}{Requests of $i$ generated at $t$.}
\nomenclature{$\alpha[t]$}{Price of energy charge at $t$.}
\nomenclature{$\beta_j$}{Price of type $j$ demand charge.}
\nomenclature{$E_{pue}$}{Power usage effectiveness.}
\nomenclature{$P_{idle}$}{Power of an idle machine.}
\nomenclature{$P_{peak}$}{Power of a fully utilized machine.}
\nomenclature{$o_{on}, o_{of}$}{Power overhead of turning a machine on and off.}
%\nomenclature{$o_{of}$}{Power overhead of turning off a machine.}
%\nomenclature{$m[0]$}{Number of switched on machines at the initial stage.}
%\nomenclature{$S[0]$}{Energy stored by the storage system at the initial stage.}
\nomenclature{$l_{in}$}{Maximum charge rate of the storage system.}
\nomenclature{$l_{out}$}{Maximum discharge rate of the storage system.}
\nomenclature{$C_s$}{Storage capacity of the storage system.}
\nomenclature{$N$}{A machine's capacity.}
\nomenclature{$\epsilon$}{Small positive number which approaches $0$.}
\nomenclature{$D_i$}{Threshold of time-shifting of workloads of $i$.}
\nomenclature{$\gamma_i$}{Reward given to $i$ per $\Psi$ requests.}
\nomenclature{$L_i$}{Revenue loss of $i$ per $\Psi$ requests.}
\nomenclature{$\rho$}{Reward factor.}
%\nomenclature{Rev}{Revenue of CDC.}
%\nomenclature{Rew}{Reward given to users.}
%\nomenclature{Wea}{wear-and-tear cost of batteries and machines.}
\nomenclature{$\zeta$}{Battery discharging cost per KWh.}
%\nomenclature{Bill}{Energy bill of CDC.}
\nomenclature{$P[t]$}{Power usage at $t$.}
\nomenclature{$P_m[t]$}{Power of machines at $t$.}
\nomenclature{$P_o[t]$}{Power usage due to turning on/off machines at $t$.}
\nomenclature{$u[t]$}{Average utilization ratio of machines at $t$.}
\nomenclature{$m[t]$}{Number of switched on machines at $t$.}
\nomenclature{$m_{on}[t]$}{Number of machines that are turned on at $t$.}
\nomenclature{$m_{of}[t]$}{Number of machines that are turned off at $t$.}
\nomenclature{$S[t]$}{Energy charged by the storage system at $t$.}
\nomenclature{$\hat{\lambda}_i[t]$}{Requests of $i$ scheduled at $t$.}
\nomenclature{$\phi_i[t]$}{Postponed requests of $i$ generated at $t$.}
\nomenclature{$\eta_i[t]$}{Postponed requests of $i$ scheduled at $t$.}
\nomenclature{$\text{Lb}$}{Lower bound of $\rho$.}
\nomenclature{$\text{Ub}$}{Upper bound of $\rho$.}
\nomenclature{$D_i^{Lb}$}{Floor of $D_i$ when $\rho=\text{Lb}$.}
\nomenclature{$w_{on}, w_{of}$}{Wear-and-tear cost of turning a machine on and off.}
%\nomenclature{$w_{of}$}{Wear-and-tear cost of turning off a machine.}
\printnomenclature

\section{Introduction}\label{sec:introduction}

To support the emergence of numerous cloud computing services,
 power-hungry data centers have collectively consumed 38 GW electricity world-wide as of 2012 (an
 increase of 63\% compared to 2011),
 placing a surging pressure on operators to optimize energy management \cite{ComputerWeekly_DatacenterPowerDemand_2012}.
Further, combined with the growing electricity prices, energy
cost has taken up nearly 15\% of data center operator's total
cost of ownership \cite{Greenberg2008}.

Consequently, as driven by the increasing popularity of real-time pricing \cite{Mohsenian-RadWongJatskevichSchoberLeon-Garcia} and peak pricing \cite{PSU_VirtualPowerPricing_HotCloud_2014},
\emph{data center demand response}, broadly interpreted as reshaping
the energy consumption of data centers, has been surfacing as a crucial approach
for saving energy cost \cite{AdamWierman_DataCenterDemandResponse_Survey_IGCC_2014}.
Towards this end, several resource management knobs, such as
workload shifting/scheduling \cite{Liu:2013:DCD:2465529.2465740,LinWiermanAndrewThereska,Le:2011:REC:2063384.2063413,PSU_PeakDemandCharge_MASCOTS_2014,ShaoleiYuxiongFei_ICDCS_2012,Xu:2014:RED:2602044.2602048}
have been extensively leveraged to shift workloads
%(and energy demand, too)
to time periods with lower
electricity prices and/or shave peak power demand to avoid high demand charges.
These knobs, albeit appealing for energy cost saving,
often result in some performance losses.
For instance, although
delay-tolerant batch workloads often do
not have as stringent delay requirements as
interactive workload,
shifting them
to time periods with low electricity prices for cost saving
still results in longer completion time and performance degradation.

While (enterprise) data centers serving internal workloads
may leverage various resource management knobs
and trade performance degradation for energy cost saving
or vice versa, public cloud data center (CDC) operators serving
cloud tenants
cannot readily do so, because workloads are managed by tenants
while CDC operators must meet the rigid service level agreements (SLAs).
Specifically, tenants in a CDC are often charged
based on their usages of CDC resources \cite{Li2010},
regardless of when they use the cloud resources.
Thus, tenants would simply submit their workloads whenever
available and complete them as soon as possible, even if their workload is delay-tolerant and
can be shifted to later times without much inconvenience.
On the other hand, due to the lack of control of tenants' workloads, CDC operators can only \emph{passively}
process the submitted workloads in
the order tenants submit them.
As a consequence, there exists a ``split-incentive'' hurdle in CDCs:
operators would like to defer some workloads
to time periods with lower electricity prices for energy
cost saving, but do not have control over tenants' workloads to enforce their wish.
%tenants manage their workloads but have no incentives
%to defer their workloads, even though doing so brings
%no or little inconvenience.

Towards breaking the ``split incentive'' hurdle,
a few recent studies \cite{PSU_PeakDemandCharge_MASCOTS_2014,Ishakian:2012:CEW:2442626.2442650,PSU_VirtualPowerPricing_HotCloud_2014,Hamed_PricingCloudTenants_Globecom2015,PSU_FairCostAttribution_Tenant_MASCOTS_2015}
have begun to investigate market-based mechanisms (e.g., dynamic pricing)
that provide tenants with monetary incentives to cooperate with
CDC operators for cost saving.
These pricing mechanisms set prices for cloud resources
dynamically, thereby
encouraging tenants to use cloud resources
at times with lower electricity prices such that the CDC operator
can reduce its energy cost, too.

While many of the existing market-based mechanisms are promising for reducing the CDC's energy
cost and/or lowering tenants' cloud costs,
they often rely on complex designs, which may be difficult to implement
in practice. For example, \cite{Ishakian:2012:CEW:2442626.2442650,PSU_VirtualPowerPricing_HotCloud_2014}
attribute each tenant's cost based on its contribution to the overall energy bill,
and this requires the CDC operator reveals to tenants sufficient information about its energy bill and each tenant's contribution.
In addition, time-dependent dynamic pricing, as often considered
 in these prior studies, result in cost uncertainties/concerns for tenants,
 who, in order to opportunistically launch their cloud
 resources for workload processing,
 have to frequently adjust their scheduling decisions
 based on real-time price information. This is particularly
 unappealing for small/medium tenants who may
 see their costs increased due to the lack
of required expertise to properly respond to dynamic pricing or other market mechanisms.

To address the aforementioned limitations of
current pricing approaches \cite{PSU_PeakDemandCharge_MASCOTS_2014,Ishakian:2012:CEW:2442626.2442650,PSU_VirtualPowerPricing_HotCloud_2014,Hamed_PricingCloudTenants_Globecom2015,PSU_FairCostAttribution_Tenant_MASCOTS_2015},
we propose a new and practical pricing policy, called Usage-based Pricing with Monetary Reward (\ouralg), which \emph{keeps pricing for cloud resources unchanged for a long period yet still being able to exploit tenants' workload scheduling flexibilities
via rewards}. Specifically, while \ouralg charges tenants based on a Usage-based Pricing (UP)
as used by today's major cloud operators, it rewards
tenants \emph{proportionally} based on the time length that tenants
set as deadlines for completing their workloads.

By using \ouralg, tenants with no delay
tolerance for their workloads are not affected,
whereas tenants who are willing to defer
their workloads for financial compensation
are given an opportunity to reveal their workload
scheduling flexibilities, i.e., by which time
their workloads need to be completed. Thus, on one hand,
\ouralg is non-intrusive to tenants, and participating
in \ouralg is fully voluntary.
On the other hand, the CDC operator can exploit
the scheduling flexibilities provided
by motivated tenants and properly schedule these
workloads for energy cost saving subject to deadline constraints.

%To summarize, the contributions
% of this paper are as follows:
%\begin{itemize}
%\item We propose \ouralg, a new and practical pricing policy
%that extracts tenants' workload scheduling flexibilities through incentives
%and allows the CDC operator to re-schedule (flexible) workloads for energy cost saving.
%\item

To derive UPMR, we will first present a detailed model for the interactions between tenants and the CDC operator, and then formulate the operator's (non-convex) problem of energy cost minimization, which is decomposed into several convex sub-problems.
%\item
%
Using real-world workload traces, we show that \ouralg can effectively reduce the operator's energy cost by 12.9\% and increase its profit  by 4.9\%, compared to state-of-the-art approaches used by today's CDC operators.
%
%\end{itemize}

\section{Related Work}\label{sec:relatedwork}
%makeindex -s nomencl.ist -o TSG.nls TSG.nlo

%In this section, we provide a snapshot of the literature
%and highlight the differences between our work and related works.

Optimizing data center energy efficiency has attracted great attention in the past years.
For example, ``power proportionality'' has been investigated in various contexts,
such as dynamically turning on/off servers for interactive workloads \cite{LinWiermanAndrewThereska,GuenterJainWilliams,Gandhi:2012:ADR:2382553.2382556},
tuning processor speed for online data-intensive services \cite{UMich_OLDI_Services}.
Other techniques include energy storage control in concert with IT resource management \cite{GovindanWangSivasubramaniamUrgaonkar_ASPLOS_2012, Neely_UPS_Sigmetrics, EnergyStorage_Sigmetrics},
thermal-aware load scheduling \cite{Tang:2007:TTS:1545007.1545141},
geographic load balancing \cite{Hamed2010, Liu:2011:GGL:1993744.1993767, RaoLiuXieLiu_2010}, renewable generation proportionality \cite{Ghamkhari2013,Mahdi_ICC_2012}, among others. More recently, there has been a growing interest in
data center demand response  \cite{DataCenterDemandResponse}.
These studies, however, all assume that data center operators have full control over the
workloads, but this assumption does not hold in CDCs where workloads are managed
by individual tenants.

%\subsection{User Incentive}\label{subsec:userincentive}
To enable demand response within CDCs,  various market-based pricing schemes have been recently
proposed to align tenants' interests with the operator's, such that
these two separate parties can cooperate towards a desired goal \cite{PSU_PeakDemandCharge_MASCOTS_2014,Ishakian:2012:CEW:2442626.2442650,PSU_VirtualPowerPricing_HotCloud_2014,Hamed_PricingCloudTenants_Globecom2015,PSU_FairCostAttribution_Tenant_MASCOTS_2015}.
More specifically, \cite{PSU_VirtualPowerPricing_HotCloud_2014,PSU_FairCostAttribution_Tenant_MASCOTS_2015}
considered fair energy cost attribution and split the cost
among tenants based on their individual contributions to the overall energy bill.
Similarly, in \cite{Ishakian:2012:CEW:2442626.2442650} the authors proposed to
fairly share the energy bill among tenants by incentivizing
 them to disclose their workload flexibility to the CDC operator.
In these studies, the CDC operator needs
to provide to tenants information about
electricity rates as well as the peak power demand, which may be difficult to implement.
Further, tenants' costs of using cloud resources may become
highly unpredictable under these pricing policies, causing
business concerns/inconveniences for tenants' budgeting.
In \cite{Hamed_PricingCloudTenants_Globecom2015}, the authors proposed dynamic pricing
for cloud resources, propagating the total energy bill onto tenants' costs.
In this case, tenants are incentivized to modulate their workloads based on the real-time price information provided by the CDC operator
and hence need to periodically adjust their workload scheduling decisions.
This represents a significant barrier, especially for small/medium tenants
who do not have the required expertise to do so.

In contrast, \ouralg is a practical mechanism that is easy to implement
and uses monetary rewards to extract tenants' workload
scheduling flexibilities, allowing the CDC operator to manage these workloads
accordingly for demand response and energy cost saving. A key advantage
of \ouralg, as compared to the existing solutions is that,
tenants' participation is fully voluntary: only tenants who are interested
in trading their workload deferments for rewards need to provide
scheduling flexibilities, while other uninterested tenants
are not affected.

%We propose UPMR in this paper. First, UPMR use monetary reward to incentivize users of CDC to allow the operator to time-shift their workloads over a proper time span. Second, UPMR is simple and easy to budget since the price of instances and reward for unit requests is pre-decided at the SLA-signing stage and will remain unchanged during a relatively long period. Third, users with UPMR do not need to participate in workload modulation, which will be done by CDC transparently to users.

\section{System Model}

In this section,  we model both cloud tenants' and the CDC operator's decisions,
based on a time-slotted system.

\subsection{Cloud Tenants}\label{sec:user}

Tenants with different  workloads/jobs naturally have different sensitivities to postponing their workloads. For example, batch workloads such as MapReduce-based tasks can
often tolerate a longer deferment without affecting tenants' normal businesses, than user-interactive workloads such as web search.
Here,
we classify tenants into $I$ types, each one having a different sensitivity with respect to workload
latency. More specifically,
we use a \emph{revenue loss} factor $\kappa_i \in \mathbb{R}^{+}_{0}$ to quantify type-$i$
 tenants' sensitivities to deferring their workloads: the more sensitive to deferment, the
 larger $\kappa_i$.
For tenants running interactive applications with no delay tolerance, we set $\kappa_i \to \infty$ and refer to these tenants
 as \emph{inelastic} tenants. All other types of tenants are referred
to as \emph{elastic} tenants for whom deferring workloads is
acceptable to a certain extent.

Suppose that type-$i$ elastic tenants' (potential) revenue losses
are nondecreasing and convex in the maximum number of time slots $D_i$ by which
 their workloads can be deferred. In practice, tenants' workloads are
 often processed prior to the maximum deferment of $D_i$ time slots, and thus
 $D_i$ can be viewed as the maximum delay tolerance for type-$i$ elastic tenants.
For example,  we consider the following
formulation to measure type-$i$ elastic tenants' revenue losses per $\Psi$ requests\footnote{Our model also applies to any other revenue loss function, provided that it is nondecreasing and convex over deferment parameter $D_i$.}:
\begin{equation}\label{equ:userloss}
L_i=\kappa_i D_i.
\end{equation}
We also use the following reward function to be offered to type-$i$ tenants per $\Psi$ requests due to their workload flexibility:
\begin{equation}\label{equ:discount}
\gamma_i=\rho \log(1+ D_i),
\end{equation}
where the reward factor $\rho$ is always non-negative
and the deferment threshold specified by tenants should satisfy
 \begin{equation}\label{con:di}
0 \leq D_i \leq D_{{max}},
\end{equation}
in which $D_{max} \in \mathbb{Z}^{+}_{0}$ is a fixed positive integer number. %denotes the maximum deferment bound. % for tenants' workloads. %This also prevents tenants from gaining unbounded rewards from the CDC operator
%by reporting an arbitrarily large $D_i$.

\vspace{0.1cm}

\textbf{Tenants' Decisions}: We assume that tenants are rational, i.e., each tenant makes decision on its deferment threshold $D_i$ for maximizing
its net benefit (reward minus revenue loss). Thus, tenant's optimization problem can be formulated
as
\begin{equation}\label{pro:user}
\begin{aligned}
& \underset{D_{i}}{\text{maximize}}
& & \gamma_i-L_i, \ \ \text{s.t., constraint }(\ref{con:di}),
%& \text{with expression}
%& & (\ref{equ:userloss}),(\ref{equ:discount})\\
\end{aligned}
\end{equation}
where the reward $\gamma_i$ and revenue loss $L_i$ are
given by \eqref{equ:discount} and \eqref{equ:userloss}, respectively.
Note that, although $D_i$ should be set as an integer value due to time-slotted
 system model, we relax it to be continuous to simplify the problem (\ref{pro:user}).
 %\footnote{Identically, we let $m_{on}[t]$, $m_{of}[t]$, $\phi_i[t]$ and $\eta_i[t]$ in the remaining of this paper  take real values instead of integers for analytical tractability and problem simplification.}
Thus, the optimization problem (\ref{pro:user}) is convex, whose solution is
\begin{equation}\label{equ:di}
D_i=\max\left(\min\left(\frac{\rho}{\kappa_i}-1,D_{max}\right),0\right).
\end{equation}

%\begin{equation}\label{equ:di}
%D_i=\max\left(\min\left(\frac{\rho}{\kappa_i}-1,D_{max}\right),0\right).
%\end{equation}

\subsection{CDC Operator}\label{sub:cdccontrol}

%\begin{figure}[!t]
%	\centering
%	\includegraphics[width=1\linewidth]{fig/framework}
%	\caption{A simple illustration of the system model in this paper: a CDC encourages its elastic users to postpone their workloads for monetary reward.}
%\vspace{-0.2cm}
%\label{fig:framework}
%\end{figure}

The CDC operator earns revenue by serving tenants' workloads
and incurs energy cost for workload processing. By using \ouralg,
the CDC operator also has a cost of rewards paid to incentivize
workload deferment by elastic tenants.

\vspace{0.1cm}

\textbf{Revenue}: Without loss of generality, we assume that each request of user will consume the same amount of CDC resources, which includes CPU, memory, storage and bandwidth, etc. Suppose $\delta \in \Re^{+}_{0}$ per Dollar denotes the price of CDC resources for processing $\Psi$ requests. $\tau$ denotes the amount of time slots over a billing cycle.
We can formulate the revenue that is earned by the CDC operator over a billing cycle as
\begin{equation}\label{equ:revenue}
\text{Revenue}=\delta \sum_{t=1}^{\tau}{ \sum_{i=1}^{I}{\frac{\lambda_i[t]}{\Psi}}},
\end{equation}
where $\lambda_i[t] \in \mathbb{Z}^{+}_{0}$ denote the number of service requests that are generated by type-$i$ tenants at time slot $t$.
%
%where $\theta_{i,k}[t] \in \mathbb{Z}^{+}_{0}$ denotes the amount of type-$k$ cloud instances rented by type-$i$ users at time slot $t$, which is decided by tenants and can be either constant or time-varying.%\footnote{Traditionally, the amount of instances rented by a user remains unchanged during an relatively long period. However, some paper suggest that users can adjust their rental of instances periodically to save cost based on their variable workloads and the real time prices of instances such as in \cite{Zhan2015}.}

 %In this paper, we assume that users decide their rental of instances based on the amount of requests generated at each time slot, which can be calculated as
%\begin{equation}\label{equ:thetai}
%\theta_i[t]=\left\lceil \frac{\lambda_i[t]}{\nu} \right\rceil, \quad \forall i,t.
%\end{equation}
%Here, $\lambda_i[t]$ denotes the requests generated by type $i$ users at time slot $t$, $\nu$ denotes the number of requests can be processed by an instance within a time slot and $\lceil \cdot \rceil$ represents the ceil function. Note that the definition of $\theta_i[t]$ as in (\ref{equ:thetai}) can be replaced by a constant if users are not able/willing to adjust their rental of instances over  time.

\vspace{0.1cm}

\textbf{Reward}: We write the total rewards over a billing cycle as
\begin{equation}\label{equ:reward}
\text{Reward}=\sum_{i=1}^{I}{\gamma_i\sum_{t=1}^{\tau}{\frac{\lambda_i[t]}{\Psi}}}.
\end{equation}
We assume that the CDC operator knows or can estimate\footnote{In practice, the CDC operator can learn tenants' loss factors from experiments or historical data, using, e.g., \emph{calibration period of pilot trials} \cite{Ha2012}.} tenants' revenue loss factors $\kappa_i$.
Thus, the CDC operator can calculate tenants' responses to its offered reward rate according
to (\ref{equ:di}). By replacing $D_i$ in (\ref{equ:discount}) with (\ref{equ:di}), we get
\begin{equation}\label{equ:discountoverrho}
\gamma_i=\rho \log\left(1+\max\left(\min\left(\frac{\rho}{\kappa_i}-1,D_{max}\right),0\right)\right).
\end{equation}

\vspace{0.1cm}

\textbf{Wear-and-Tear Cost}: The CDC operator incurs a  wear-and-tear cost during operation, e.g., during server on/off and battery charge/discharge cycles, which can be formulated as
\begin{equation}\label{equ:battarycost}
\begin{aligned}
\text{Wear}= \; &\zeta\sum_{t=1}^{\tau}{\max\{-S[t],0\}}\\
&+w_{on}\sum_{t=1}^{\tau}m_{on}[t]+w_{of}\sum_{t=1}^{\tau}m_{of}[t],
\end{aligned}
\end{equation}
where $\zeta \in \Re^{+}_{0}$ ($\$/$KWh) measures the estimated battery wear-and-tear cost, $S[t]$ (KWh) denotes the energy
 consumption for charging the storage system at time slot $t$, and $S[t]< 0$ means that the storage system discharges energy to  power up the CDC as a supplement to grid power.
As for parameters $m_{on}[t]$ and $m_{of}[t]$, they denote the number of physical machines that are turned on/off at time slot $t$, respectively.
In addition, $w_{on} \in \Re^{+}_{0}$ ($\$$) and $w_{of} \in \Re^{+}_{0}$ ($\$$) measure the wear-and-tear cost of machine due to turning on/off machine, respectively.

\vspace{0.1cm}

\textbf{Energy Cost}: Typically, the CDC operator is charged for electricity by the utility based on two parts: \emph{energy charge} and \emph{demand charge}. Here, energy charge is calculated based on the amount of energy consumption,
while demand charge is calculated according to the peak demand, e.g., measured over each 15 minutes interval within a billing cycle \cite{Oudalov2007}. Thus, the CDC's energy cost over a billing cycle can be calculated
as
\begin{equation}\label{equ:bill}
\text{Bill}=\sum_{t=1}^{\tau}{\alpha[t]P[t]}+\sum_{j=1}^{J}\frac{\beta_j}{T} \max_{t \in A_j}{P[t]},
\end{equation}
where $T$ denotes the length
of each time slot (e.g., hour), $P[t] \in \Re^{+}_{0}$ (KWh) denotes the energy usage at time slot $t$, $\alpha[t] \in \Re$ ($\$$ per KWh) denotes the energy price at time slot $t$,
$\beta_j \in \Re^{+}_{0}$ ($\$$ per KW) denotes the price of type-$j$ demand charge,\footnote{Utilities may impose multiple demand charges for different time intervals, e.g., daytime/night or winter/summer demand charge \cite{PGE}.}
          and $A_{j}$ denotes the set of time slots falling into
          the type-$j$ demand charge
          window. The considered energy billing model is quite general
and includes Time-Dependent Pricing (TDP) as a special case, where
the CDC is charged only for energy consumption based on time-dependent prices. %(i.e., $\beta_j=0$ and $\exists t, t^{'} \in \{1,\cdots,\tau\}$, $\alpha[t] \ne \alpha[t^{'}]$).

\vspace{0.1cm}

\textbf{Energy consumption.}\label{subsec:power} The total energy usage of the CDC can be calculated at each time slot by taking into consideration the server energy consumption, non-IT energy (captured by power usage effectiveness, i.e., the ratio of total
data center energy to IT energy), overheads due to turning machines on/off and the storage capacity of CDC. Thus, we can write
\begin{equation}\label{equ:energyusage}
P[t]=\max\left(E_{pue}(P_m[t]T+P_o[t])+S[t],0\right), \quad \forall t,
\end{equation}
where $E_{pue} \in [1,\infty)$ denotes power usage effectiveness, %(which
%can be estimated based on environmental factors such as outside
%air temperature),
$P_m[t] \in \Re^{+}_{0}$ (KW)  denotes the power of \emph{active} machines that are being used,
  and $P_o[t] \in \Re^{+}_{0}$ (KWh) is the energy overhead for
 turning on/off machines at time slot $t$.
As in \cite{Li2015}, the power of active machines is
expressed as
\begin{equation}\label{equ:pm}
P_m[t]=m[t]\left(P_{idle}+(P_{peak}-P_{idle})u[t]\right), \quad \forall t,
\end{equation}
where $m[t]  \in \Re^{+}_{0}$ denotes the number of active machines, $u[t] \in [0,1]$ denotes the average server utilization, $P_{idle} \in \Re^{+}_{0}$ (KW) and $P_{peak}\in [P_{idle},\infty)$ (KW) denote the energy usage of a machine in idle and fully utilized conditions, respectively.  Next, we calculate the amount of active machines by
\begin{equation}\label{equ:mt}
m[t]=\sum_{t^{'}=1}^{t} (m_{on}[t^{'}]-m_{of}[t^{'}])+m[0], \quad \forall t,
\end{equation}
where $m[0] \in \mathbb{Z}^{+}_{0}$ is the number of switched on machines at time $0$. We formulate the average server utilization as
\begin{equation}\label{equ:utilizationratio}
u[t]=\frac{\sum_{i=1}^{I}{\hat{\lambda}_i[t]}}{N m[t]}, \quad \forall t,
\end{equation}
where $\hat{\lambda}_i[t] \in \Re^{+}_{0}$ is the service requests of type-$i$ tenants scheduled at time slot $t$ and $N \in \mathbb{Z}^{+}_{0}$  is the (average) number of service requests that can be hosted by a machine in a time slot.
Here, the number of \emph{scheduled} requests is modeled as
\begin{equation}\label{equ:hatlambda}
\hat{\lambda}_i[t]=\lambda_i[t]-\phi_i[t]+\eta_i[t], \quad \forall i,t,
\end{equation}
where $\phi_i[t]$ and $\eta_i[t]$ denote the deferred requests of type-$i$ tenants generated and scheduled at time slot $t$, respectively. %That is, requests scheduled at time $t$ equals to the amount of requests generated and scheduled at time slot $t$ (i.e., $\lambda_i[t]-\phi_i[t]$) plus the amount of deferred requests which are scheduled at time slot $t$ (i.e., $\eta_i[t]$).

The power usage of turning machines on/off is modeled as
\begin{equation}\label{equ:po}
P_o[t]=o_{on}m_{on}[t]+o_{of}m_{of}[t], \quad \forall t,
\end{equation}
where $o_{on} \in \Re^{+}_{0}$ (KWh) and $o_{of}  \in \Re^{+}_{0}$ (KWh) denote the energy consumption of turning on/off a machine, respectively.

\vspace{0.1cm}

\textbf{Operational Constraints}: Next, we list the constraints that the CDC operator faces when it makes control decisions.

First, the reward factor and the number of machines that are turned on/off at each time slot
must be non-negative:
\begin{equation}\label{con:rholowerbound}
\rho \geq 0,
\end{equation}
\begin{equation}\label{con:mon_and_off}
m_{on}[t] \geq 0, \ m_{of}[t] \geq 0, \quad \forall t.
\end{equation}
%\begin{equation}\label{con:moff}
%m_{of}[t] \geq 0, \quad \forall t.
%\end{equation}

Second, the number of active machines should be large enough to process requests scheduled at each time slot:
\begin{equation}\label{con:machine}
\begin{aligned}
&\sum_{t^{'}=1}^{t} (m_{on}[t^{'}]-m_{of}[t^{'}])+m[0] \geq\\
 &\frac{\sum_{i=1}^{I}(\lambda_i[t]-\phi_i[t]+\eta_i[t])}{ N} , \quad \forall t,
\end{aligned}
\end{equation}
which ensures that the average utilization as shown in (\ref{equ:utilizationratio}) is always no greater than $1$.

Third, if the CDC is equipped with an on-site energy storage unit, then the energy storage unit's charge and discharge rates should be limited by the charger inverter's ratings:
\begin{equation}\label{con:storagecharge}
-l_{out}T\leq S[t] \leq l_{in}T, \quad \forall t,
\end{equation}
where $l_{out} \in \Re^{+}_{0}$ (KW) and $l_{in} \in \Re^{+}_{0}$ (KW) denote the
 energy storage unit's maximum discharge and charge rates, respectively. The amount of energy stored in the storage unit should be non-negative and not exceed its storage capacity:
\begin{equation}\label{con:storagelimitation}
0\leq \sum_{t^{'}=1}^{t} S[t^{'}] +S[0]\leq C_s, \quad \forall t,
\end{equation}
where $S[0]\in \Re^{+}_{0}$ (KWh) is the stored energy at
 time $0$ and $C_s \in \Re^{+}_{0}$ (KWh) is the storage capacity.

Fourth,  $\phi_i[t]$ and $\eta_i[t]$ satisfy
\begin{equation}\label{con:philowerbound}
0 \leq \phi_{i}[t] \leq \lambda_i[t], \quad \forall i,t,
\end{equation}
\begin{equation}\label{con:etalowerbound}
\eta_{i}[t] \geq 0, \quad \forall i,t.
\end{equation}
That is, the amount of deferred requests generated at time slot $t$ must be non-negative and
 no more than the amount of requests newly generated at time slot $t$. Also, the amount of deferred requests scheduled at time slot $t$ must be non-negative.

Fifth, considering that the amount of requests generated from the beginning of a billing cycle up to any time slot must be no less than the amount of requests that are scheduled within these periods, we must also have
\begin{equation}\label{con:phigeqeta1}
\sum_{t^{'}=1}^{t} {(\lambda_i[t^{'}]-\phi_i[t^{'}]+\eta_i[t^{'}])} \leq \sum_{t^{'}=1}^{t} {\lambda_i[t^{'}]}, \quad \forall i,t,
\end{equation}
which can be simplified to be as
\begin{equation}\label{con:phigeqeta}
\sum_{t^{'}=1}^{t} {(-\phi_i[t^{'}]+\eta_i[t^{'}])} \leq 0, \quad \forall i,t.
\end{equation}

Last but not least, to ensure that the requests of type-$i$ tenants generated from the beginning
up to the time slot $t$ should be scheduled no later than time slot $t+\lfloor D_i \rfloor$, where $\lfloor \cdot \rfloor$ represents a floor function, we have
\begin{equation}\label{con:hatgeqnohat1}
\sum_{t^{'}=1}^{t+\lfloor D_i \rfloor} {(\lambda_i[t^{'}]-\phi_i[t^{'}]+\eta_i[t^{'}])} \geq \sum_{t^{'}=1}^{t} {\lambda_i[t^{'}]}, \quad \forall i,t,
\end{equation}
which can be simplified to be as
\begin{equation}\label{con:hatgeqnohat}
\sum_{t^{'}=t+1}^{t+\lfloor D_i \rfloor} {\lambda_i[t^{'}]} +\sum_{t^{'}=1}^{t+\lfloor D_i \rfloor} {(-\phi_i[t^{'}]+\eta_i[t^{'}])} \geq 0, \quad \forall i,t.
\end{equation}

That is, the CDC cannot defer tenants' requests beyond their deferment thresholds. Note that, $\forall t>\tau$,
$\lambda_i[t]=0$, $\phi_i[t]=0$ and $\eta_i[t]\geq 0$,  which indicates that
the CDC may schedule some of tenants' requests at time slot in the next billing cycle without violation of the thresholds specified by the tenants.

\section{\ouralg Pricing Algorithm: A Profit \\ Maximization Approach}

%In this section, we present a decomposition method
%to maximize the CDC operator's profit. % using the proposed \ouralg mechanism.

\subsection{Problem Formulation}\label{subsec:profitmaximization}
We first formulate the CDC operator's profit over a billing cycle as its revenue minus its various expenses:
\begin{equation}\label{equ:profit}
\text{Profit} = \text{Revenue}-\text{Reward}-\text{Wear}-\text{Bill},
\end{equation}
where Revenue, Reward, Wear and Bill are defined by (\ref{equ:revenue})-(\ref{equ:po}).
In this paper, we assume that the CDC operator optimizes its decisions at the beginning of each billing cycle with the assumption that tenants demands can be perfectly predicted. We leave online optimization under demand uncertainty as future work.
Here, we seek to solve the following problem:
\begin{equation}\label{pro:profit}
\begin{aligned}
& \underset{ \begin{subarray}{c} \rho,m_{on}[t],m_{of}[t], \\ S[t],\phi_i[t],\eta_i[t] \end{subarray}  }{\text{Maximize}}
& & (\ref{equ:profit}) \\
& \text{Subject to}
& & (\ref{con:rholowerbound})-(\ref{con:etalowerbound}), (\ref{con:phigeqeta}), (\ref{con:hatgeqnohat}). \\
\end{aligned}
\end{equation}
Problem (\ref{pro:profit}) is not convex caused by the definition of the reward rate (\ref{equ:discountoverrho}) of objective function of (\ref{pro:profit}) and constraint (\ref{con:hatgeqnohat}) and hence
difficult to solve. %In what follows, we propose a divide method
%to solve (\ref{pro:profit}) with a low complexity.

\subsection{Decomposition Method}

Next, we propose a decomposition method to tackle the difficulty in solving (\ref{pro:profit}). %, we propose a decomposition method to solve it efficiently.
Specifically, we decompose problem (\ref{pro:profit}) into several sub-problems, each corresponding to a sub-domain of $\rho$. The details are presented
 in Algorithm \ref{alg:solution}.
\begin{algorithm}[!t]
\caption{The Decomposition Method}
\label{alg:solution}
\begin{algorithmic}[1]
\REQUIRE ~~\\
$r$ \quad \quad \quad \ \ \ \ (Loop times)\\
Solution($r$) \ (Optimal solution of problem (\ref{equ:subproblem1}) at loop $r$)\\
%\ENSURE ~~\\
%FAILURE or $totalcost$ for embedding;\\
%$G^s$:Updated residual resource;\\
%$\pi$ :embedding map;\\
\STATE $r=1$.
\STATE $\text{Lb}=0$.
\WHILE {1}
    \STATE Calculate upper bound of $\rho$ via (\ref{equ:dlb})-(\ref{equ:eee}).
    \STATE Obtain a sub-domain of $\rho$ as in (\ref{con:newrholowerbound}).
    \STATE Build the corresponding sub-optimization problem (\ref{equ:subproblem1}).
    \STATE Solve problem (\ref{equ:subproblem1}).
    \STATE Save Solution($r$).
    \IF {Ub$=\infty$}
        \RETURN $\max_{\text{r}} \{\text{Solution}(r)\}$.
    \ENDIF
    \STATE Lb$=$Ub.
    \STATE $r=r+1$.
\ENDWHILE
\end{algorithmic}
\end{algorithm}

We divide problem (\ref{pro:profit}) iteratively as follow:
\begin{itemize}
\item
\textbf{Step 1}: Initializing the lower bound of sub-domain of $\rho$, i.e., Lb, by letting $\text{Lb}=0$ based on constraint (\ref{con:rholowerbound}).
\item
\textbf{Step 2}: From (\ref{equ:di}), calculate $\lfloor D_i \rfloor$ in constraint (\ref{con:hatgeqnohat}) corresponding to Lb, i.e., for each $i$, set
\begin{equation}\label{equ:dlb}
D_i^{Lb}= \left\lfloor \max\left(\min\left(\frac{\text{Lb}}{\kappa_i}-1,D_{max}\right),0\right) \right\rfloor.
\end{equation}
\item
\textbf{Step 3}: Setting the upper bound of sub-domain of $\rho$ as
\begin{equation}\label{equ:dub}
\text{Ub}=
\begin{cases}
\infty, &\mbox{If $\vartheta=\emptyset$,}\\
\min_{i \in \vartheta}\{(D_i^{Lb}+2)\kappa_i\}, &\mbox{Otherwise}.\\
\end{cases}
\end{equation}
Here, the domain $\vartheta$ is defined as
\begin{equation}\label{equ:eee}
\vartheta=\{i\in \{1,\cdots,I\} | \text{Lb}< (D_{max}+1)\kappa_i\}.
\end{equation}
\item
\textbf{Step 4}: Obtaining a new sub-domain of $\rho$ as defined by
\begin{equation}\label{con:newrholowerbound}
\text{Lb} \leq \rho \leq \text{Ub}-\epsilon,
\end{equation}
where $\epsilon$ denotes a small positive number approaches $0$.
Then, replacing $\lfloor D_i \rfloor$ in (\ref{con:hatgeqnohat}) by constant $D_i^{Lb}$ that can be calculated by (\ref{equ:dlb}) and obtain a new affine constraint
\begin{equation}\label{con:newhatgeqnohat}
\sum_{t^{'}=t+1}^{t+D_i^{Lb}} {\lambda_i[t^{'}]} +\sum_{t^{'}=1}^{t+D_i^{Lb}} {(-\phi_i[t^{'}]+\eta_i[t^{'}])} \geq 0, \quad \forall i,t.
\end{equation}
Next, build the corresponding sub-optimization problem
\begin{equation}\label{equ:subproblem1}
\begin{aligned}
& \underset{ \begin{subarray}{c} \rho,m_{on}[t],m_{of}[t], \\ S[t],\phi_i[t],\eta_i[t] \end{subarray}  }{\text{Maximize}}
& & (\ref{equ:profit})\\
& \text{Subject to}
& & (\ref{con:mon_and_off})-(\ref{con:etalowerbound}), (\ref{con:phigeqeta}),
(\ref{equ:dlb})-(\ref{con:newhatgeqnohat}). \\
\end{aligned}
\end{equation}
\item
\textbf{Step 5}: Problem (\ref{equ:subproblem1}) is not convex. However, the optimum value of optimization variable $\rho$ in problem (\ref{equ:subproblem1}) is $\rho=\text{Lb}$. It is straightforward to see that, substitution of $\rho=\text{Lb}$ in the definition of reward rate (\ref{equ:discountoverrho}) of objective function of problem (\ref{equ:subproblem1}) gives a convex program  that can be solved via convex tools such as CVX [33].
 %Considering that problem (\ref{equ:subproblem1}) is not convex, replace it by its equivalent convex program as
%\begin{equation}\label{equ:subproblem}
%\begin{aligned}
%& \underset{ \begin{subarray}{c} m_{on}[t],m_{of}[t] \\ S[t],\phi_i[t],\eta_i[t] \end{subarray}  }{\text{Maximize}}
%& & (\ref{equ:profit})\\
%& \text{Subject to}
%& & (\ref{con:mon_and_off})-(\ref{con:phigeqeta}), (\ref{equ:dlb}), (\ref{con:newhatgeqnohat}), \\
%\end{aligned}
%\end{equation}
%where $\rho$ is substituted by a constant Lb. Then, solve (\ref{equ:subproblem}) via convex tools such as CVX \cite{CVX}.
\item
\textbf{Step 6}: If Ub $< \infty$, update Lb by letting Lb$=$Ub and go to Step 2 to traverse another sub-problem corresponding to another sub-domain of $\rho$. If Ub $= \infty$, we have traversed all sub-domain of $\rho$, i.e., all possible sub-optimization problems. Thus, we can find the global optimal solution of the original problem (\ref{pro:profit}) among the optimal solutions of these sub-optimization problems.
\end{itemize}

\vspace{.1cm}
\begin{theorem}\label{the:unique}
$\lfloor D_i \rfloor=D_i^{Lb}$, $\forall \rho \in [\text{Lb},\text{Ub}-\epsilon]$, $\forall i$.
\end{theorem}
\vspace{.1cm}

The proof of Theorem \ref{the:unique} is given in Appendix \ref{proo:unique}. From Theorem \ref{the:unique}, in each loop in Algorithm \ref{alg:solution}, we can replace $\lfloor D_{i} \rfloor$ in constraint (\ref{con:hatgeqnohat}) by $D_i^{Lb}$ and build an affine constraint (\ref{con:newhatgeqnohat}) at step $4$.

\vspace{.1cm}
\begin{theorem}\label{the:rho}
The optimal value of optimization variable $\rho$ in problem (\ref{equ:subproblem1}) equals Lb.
\end{theorem}
\vspace{.1cm}

The proof of Theorem \ref{the:rho} can be found in Appendix \ref{proo:rho}.
 From Theorem \ref{the:rho}, we can substitute $\rho$ in problem (\ref{equ:subproblem1}) by Lb and build a convex program at step $5$ in Algorithm \ref{alg:solution}.

\begin{figure}[t]
\centering
\subfigure[]
{	\label{fig:workload}\centering
	\includegraphics[width=0.46\linewidth]{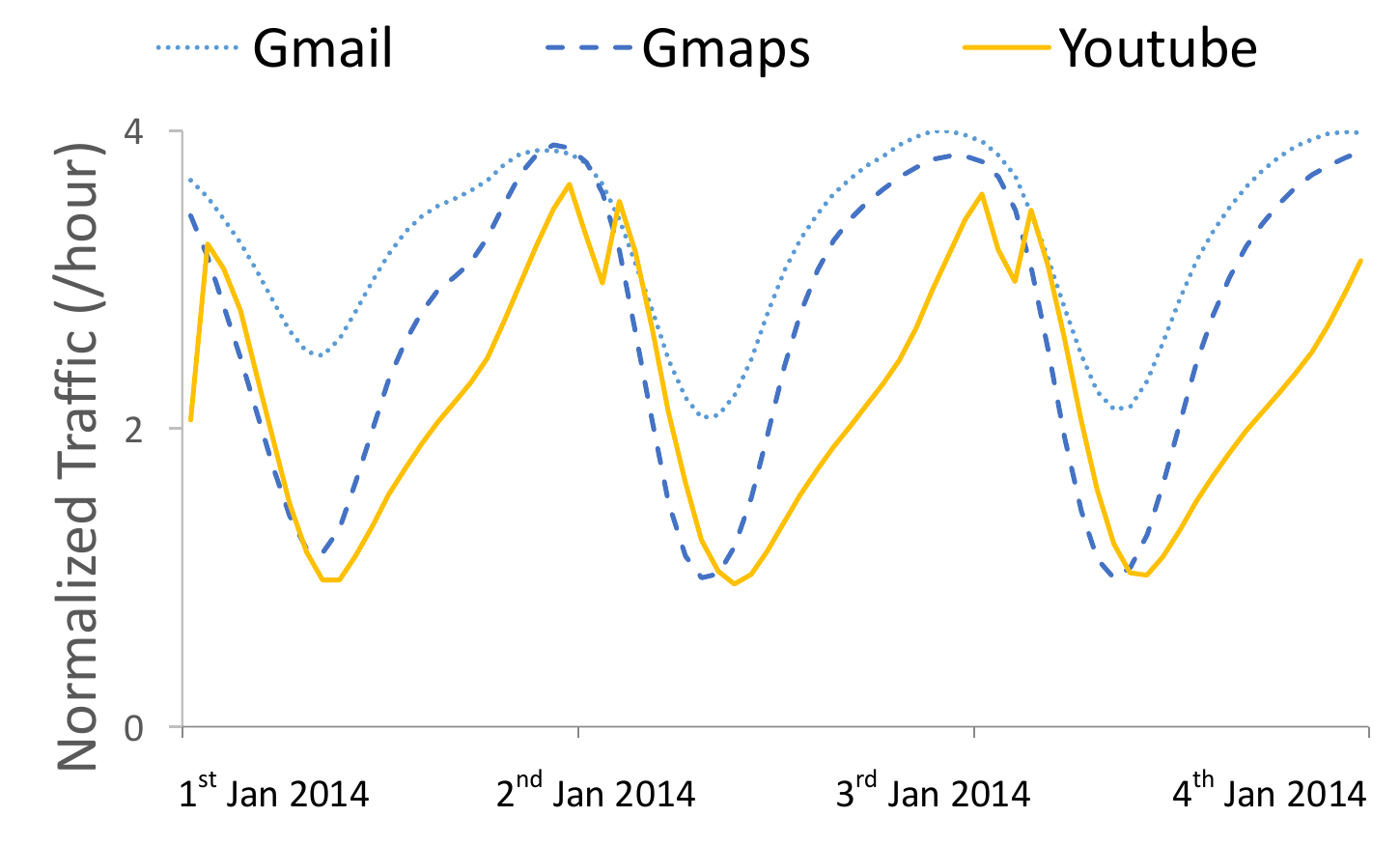}
}
\subfigure[]
{	\label{fig:energyprice}\centering
	\includegraphics[width=0.46\linewidth]{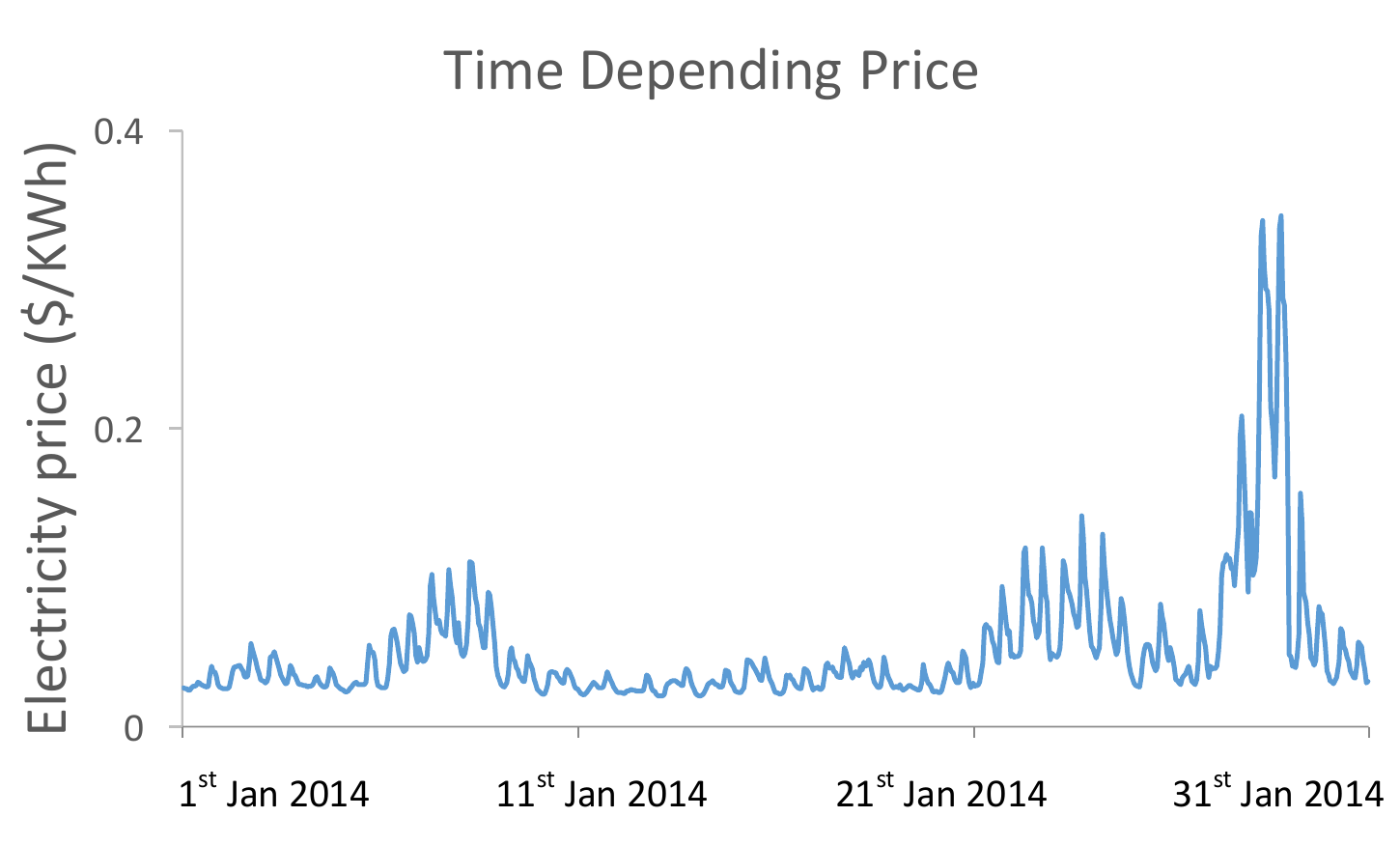}
}
\caption{Examples for data traces that we used in this paper: a) workload  traces \cite{Googledata} b) energy price of Ameren \cite{amerenrealtimeprice}.}\label{fig:data}
\end{figure}
%\begin{figure*}[t]
%\centering
%\subfigure[]
%{	\label{fig:youtube}\centering
%	\includegraphics[width=0.3\linewidth]{fig/youtube}
%}
%\subfigure[]
%{	\label{fig:gmaps}\centering
%	\includegraphics[width=0.3\linewidth]{fig/gmaps}
%}
%\subfigure[]
%{	\label{fig:gmail}\centering
%	\includegraphics[width=0.3\linewidth]{fig/gmail}
%}
%\caption{Examples for workload traces that we used in this paper: a) Fraction of Youtube US traffic, normalized \cite{Googledata}, b) Fraction of Gmaps US traffic, normalized \cite{Googledata}, c) Fraction of Gmail US traffic, normalized \cite{Googledata}.}\label{fig:workload}
%\end{figure*}
\begin{figure*}[t]
\centering
\subfigure[]
{	\label{fig:peakprofit}\centering
	\includegraphics[width=0.3\linewidth]{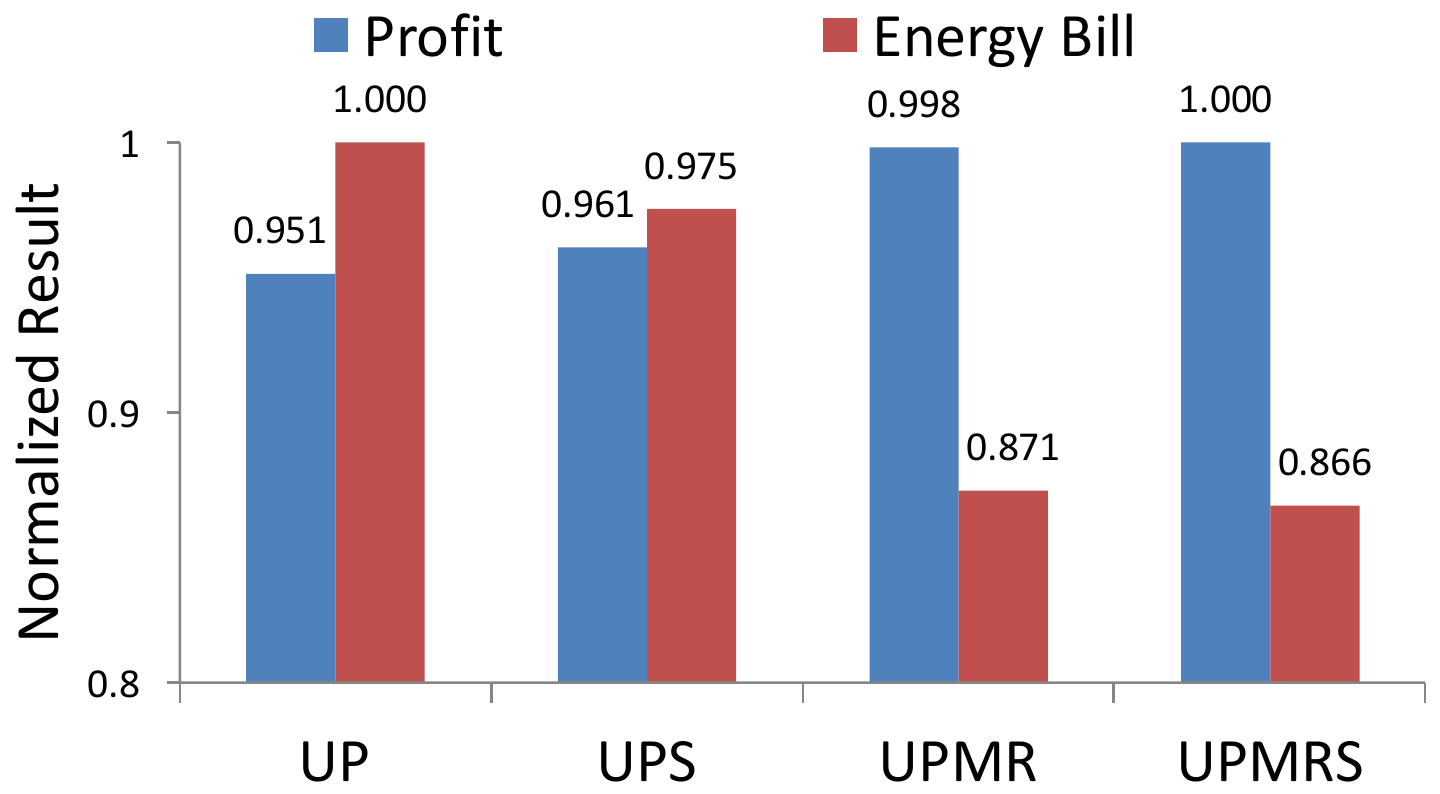}
}
\subfigure[]
{	\label{fig:peakprofitenergy1}\centering
	\includegraphics[width=0.3\linewidth]{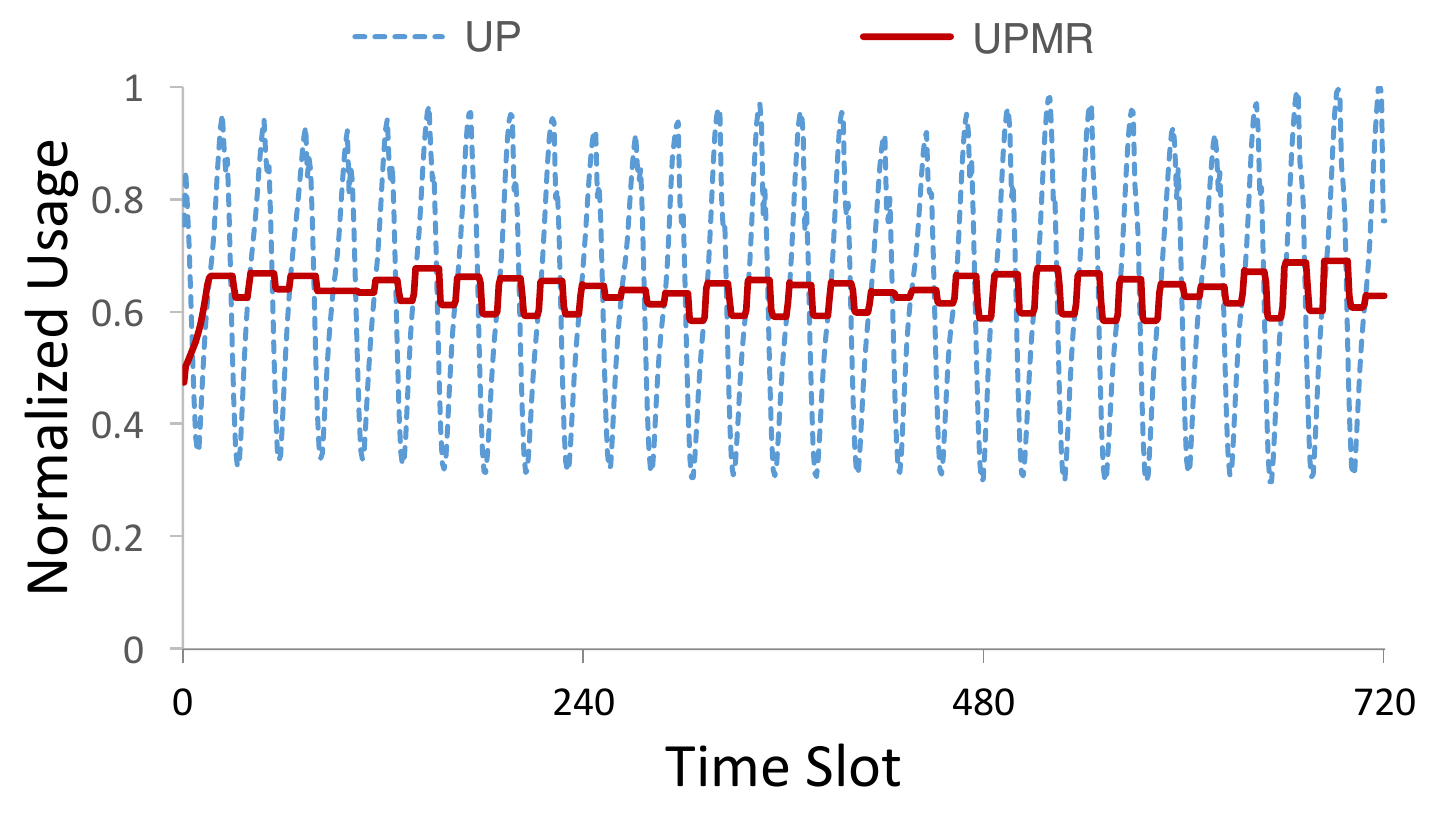}
}
\subfigure[]
{	\label{fig:peakprofitenergy2}\centering
	\includegraphics[width=0.3\linewidth]{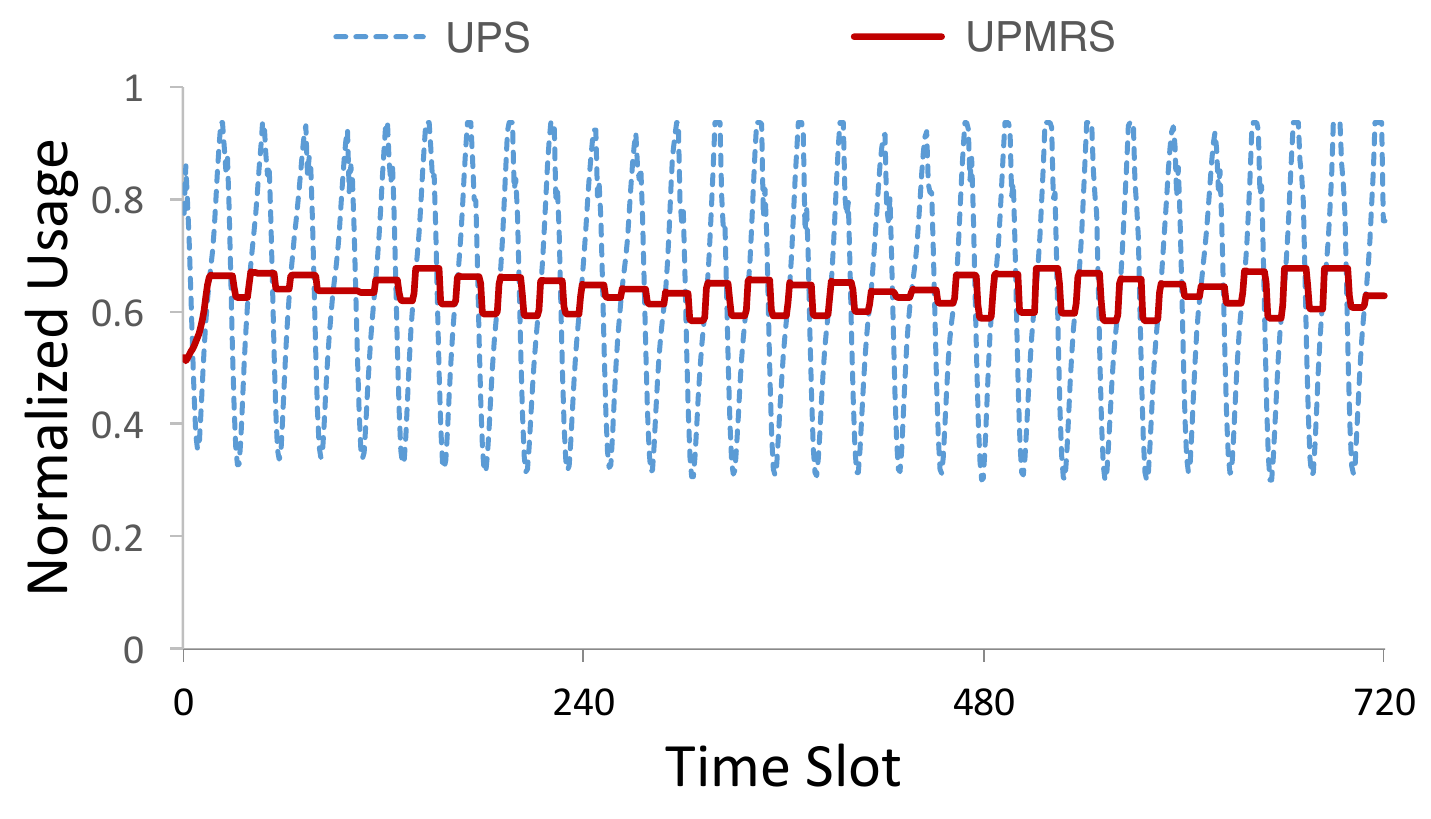}
}
\caption{Comparison among UP, UPS, UPMR and UPMRS under peak pricing: a) Normalized profit gain and energy bill of CDC, b) Energy usage of UP and UPMR, c) Energy usage of UPS and UPMRS.}\label{fig:peakprofitall}
\end{figure*}

%\begin{figure*}[t]
%\centering
%\subfigure[]
%{	\label{fig:peakwelfare}\centering
%	\includegraphics[width=0.3\linewidth]{fig/peakwelfare}
%}
%\subfigure[]
%{	\label{fig:peakwelfareenergy1}\centering
%	\includegraphics[width=0.3\linewidth]{fig/peakwelfareenergy1}
%}
%\subfigure[]
%{	\label{fig:peakwelfareenergy2}\centering
%	\includegraphics[width=0.3\linewidth]{fig/peakwelfareenergy2}
%}
%\caption{Performance comparison among FP, FPS, UPMR and UPMRS under social welfare maximization and peak pricing: a) Normalized social welfare loss and energy bill of CDC, b) Energy usage of FP and UPMR, c) Energy usage of FPS and UPMRS.}\label{fig:peakwelfareenergyall}
%\end{figure*}

%\begin{figure}[!t]
%	\centering
%	\includegraphics[width=0.7\linewidth]{fig/energyprice}
%	\caption{Examples for energy price of Ameren that we used in this paper \cite{amerenrealtimeprice}.}\label{fig:energyprice}
%\end{figure}
\begin{figure*}[t]
\centering
\subfigure[]
{	\label{fig:tdpprofit}\centering
	\includegraphics[width=0.3\linewidth]{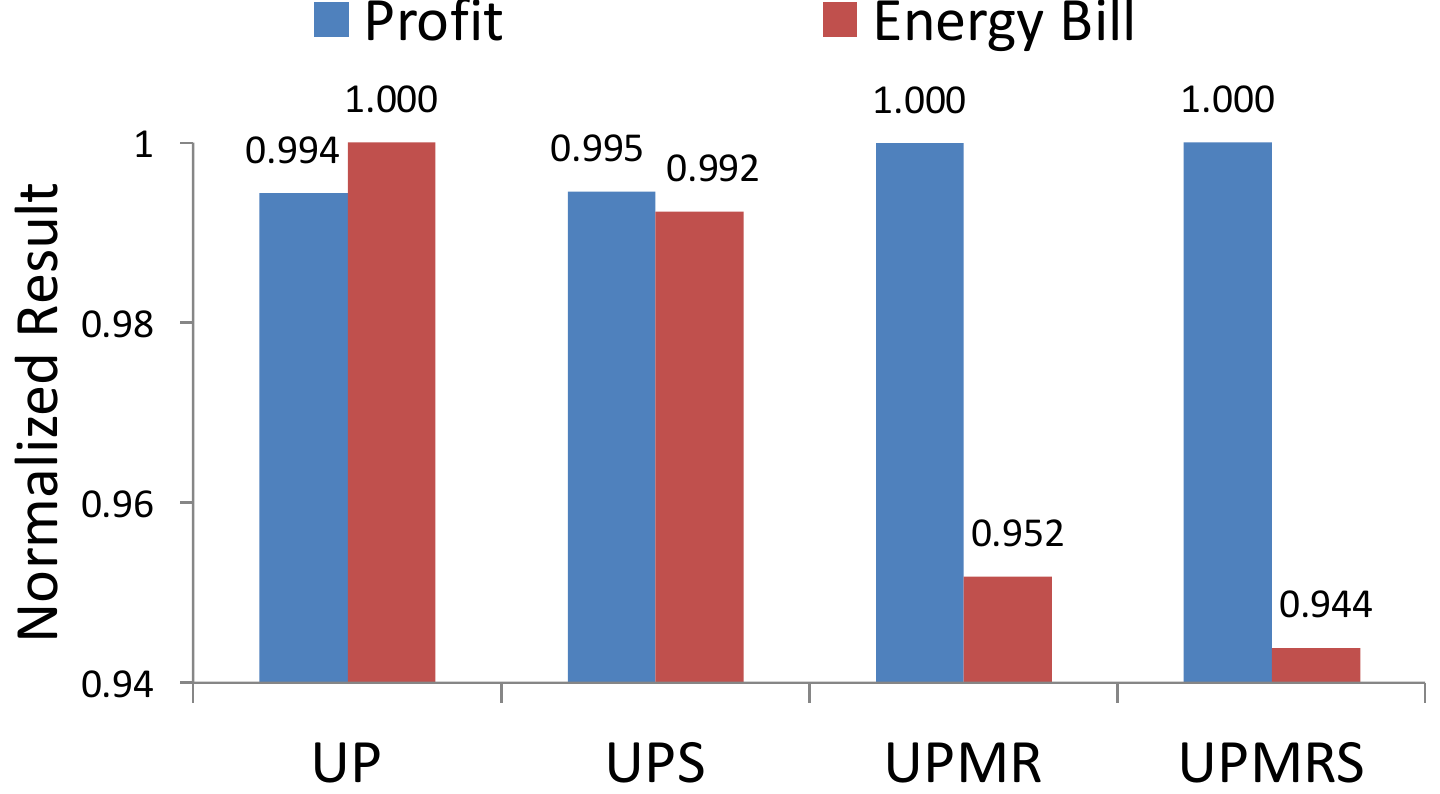}
}
\subfigure[]
{	\label{fig:tdpprofitenergy1}\centering
	\includegraphics[width=0.3\linewidth]{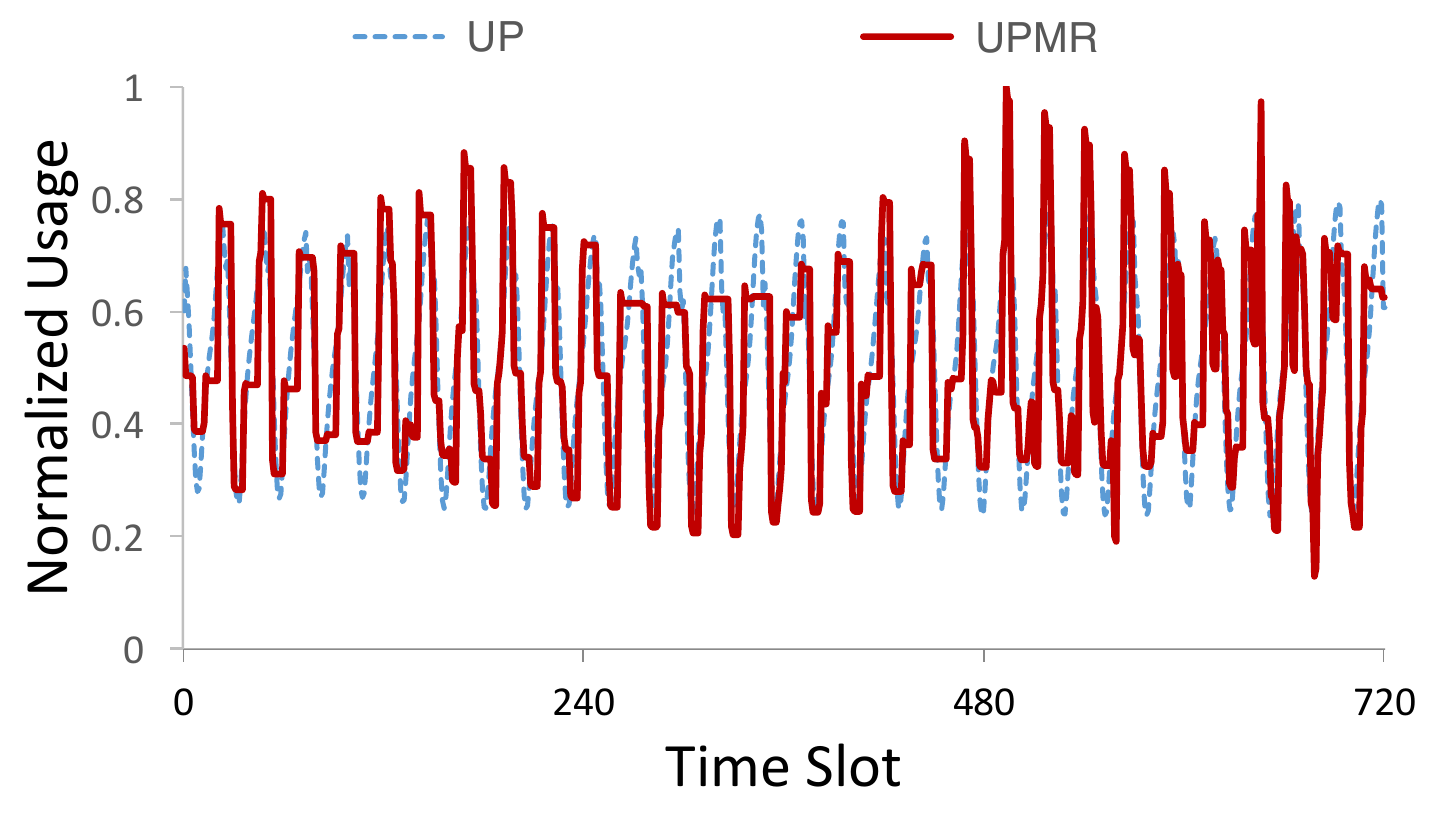}
}
\subfigure[]
{	\label{fig:tdpprofitenergy2}\centering
	\includegraphics[width=0.3\linewidth]{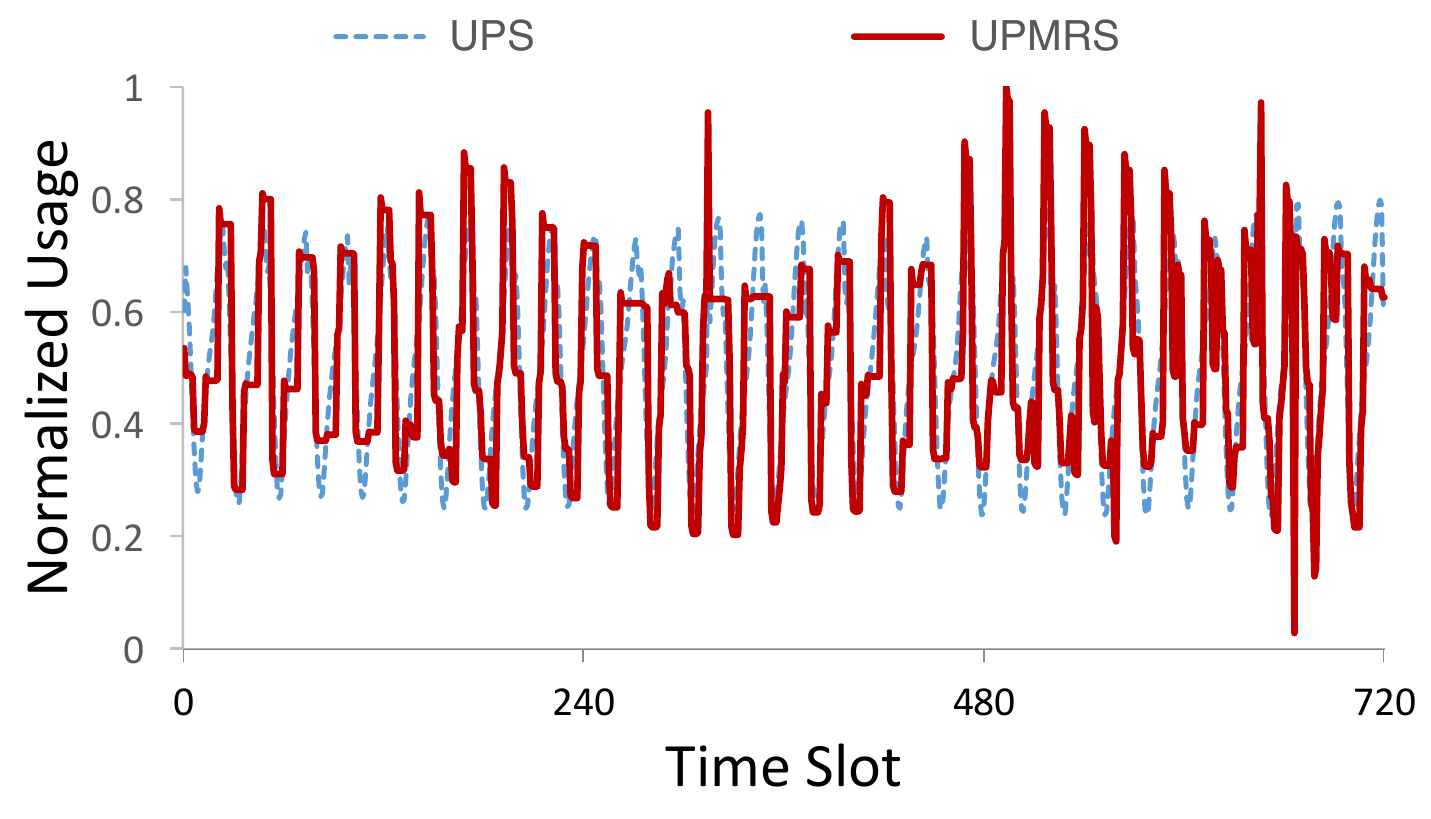}
}
\caption{Comparison among UP, UPS, UPMR and UPMRS under TDP: a) Normalized profit gain and energy bill of CDC, b) Energy usage of UP and UPMR, c) Energy usage of UPS and UPMRS.}\label{fig:tdpprofitall}
\end{figure*}

\begin{table}[!t]\centering\caption{Simulation Cases}\label{tab:cases}
\centering
\begin{tabular}{c|c|c|c}
\hline
Name & Pricing Policy & Demand Delaying & Storage Capability \\
\hline
UP & UP & No & No \\
\hline
UPS & UP & No & Yes \\
\hline
UPMR & UPMR & Yes & No \\
\hline
UPMRS & UPMR & Yes & Yes \\
\hline
\end{tabular}
\end{table}
\section{Case Study}

%Up to this point, all our analysis is theoretical, and it still remains to show the effectiveness of \ouralg in a realistic scenario for maximizing the CDC's profit. Thus, we now evaluate UPMR by conducting simulations modeling a large CDC based on real-world traces.

\subsection{Setup}
We consider a CDC serving three types of tenants: type-$1$ tenants are delay-sensitive (i.e., inelastic) while type-$2$ and type-$3$ tenants are elastic. The demands of delay-sensitive users are
 based on Youtube U.S. traffic from January 1, 2014 to January 31, 2014 \cite{Googledata}. The demands of type $2$ and $3$ users are constructed based on GMaps and GMail U.S. traffic from January 1, 2014 to January 31, 2014, respectively \cite{Googledata}. Figure~\ref{fig:workload} shows a snapshot of these workload traces that are used in the simulation. Here, elastic workloads constitute
  approximately 50\% of the overall workloads.

Let the billing cycle be 30 days and $T=1$ hour. Therefore, $\tau=30 \times 24=720$. Let $\Psi=10000$, $D_{max}=24$, $\kappa_1=\infty$, $\kappa_2=0.1$ and $\kappa_3=0.11$. Let $N=20$, $E_{pue}=1.2$, $P_{idle}=0.1$ KW, $P_{peak}=0.2$ KW, $o_{on}=0.02$ KWh and $o_{of}=0.01$ KWh. Let $m[0]=0$ and $S[0]=0$ KWh. Let $w_{on}=\$0.003$ and $w_{of}=\$0.002$ as in \cite{Qian2011}.

In addition,  we set the
CDC's energy storage capacity $C_s$, such that it can power up CDC entirely at its peak power consumption up to $30$ minutes. Both charge and discharge rates are set to be $C_s$ KW. Let $\zeta=\$0.32$ per KWh, corresponding to a 1Ah lead-acid battery at 12V as mentioned in \cite{Kontorinis2012}, which costs $2 \$$ and can stand 1300 recharge cycles at $40\%$ of depth of discharge. In this paper, we evaluate \ouralg under four cases, which are illustrated in Table \ref{tab:cases}.
%FP is a representative of a classic scenario of today's existing CDCs, who employ FP to charge their users and do not use the storage capacity of CDC to do power management. Specifically, when the pricing policy is UPMR, i.e., CDC provides users with monetary reward to incentivize them grant time-shifting of their workloads, elastic users grant time-shift their workloads over proper time spans, where the time span $D_i$ is calculated based on problem (\ref{pro:user}). Meanwhile, CDC decides the reward factor $\rho$ via Algorithm \ref{alg:solution}. Besides, for these cases where the storage capability is on, according to \cite{Urgaonkar2011},

\subsection{Under Peak Pricing}

We now evaluate \ouralg under a peak pricing scheme.
In particular,
the CDC's energy cost includes both energy charge and demand charge, where $\alpha[t]=\$0.05207$ per KWh and $\beta_j=\$ 15.59$ per KW based on electric rates of industrial power service offered by South Carolina Electric and Gas \cite{SCEG}.

Figure \ref{fig:peakprofit} shows the
CDC operator's normalized profit gain and energy cost  with respect to the maximum profit gain and energy bill among the four cases, respectively. By utilizing energy storage, the profit gain can be enhanced by $1.1\%$, while the energy bill can be reduced by $2.5\%$ when comparing with UP. Further, by employing \ouralg without utilizing the energy storage system, the profit gain can be increased by $4.9\%$ and the energy bill can be reduced by $12.9\%$ when comparing with UP. Thus, \ouralg can improve CDC's profit gain and reduce the energy bill more significantly when
 energy storage is not utilized than when it is. Additionally, \ouralg can obtain nearly the same profit gain and energy reduction as UPMRS, indicating that CDCs employing \ouralg do not need to utilize energy storage for power management.

Figure \ref{fig:peakprofitenergy1} and \ref{fig:peakprofitenergy2} show the detail of normalized energy usage of UP, UPS, UPMR and UPMRS, with respect to the maximum energy usage. We see that UPMR can effectively smooth the CDC's energy consumption. By comparing UPMR and UPMRS, we see that utilizing energy storage can also further but slightly smooth CDC's energy consumption.

%Further, we assume that the CDC is a social planner as mentioned in Section \ref{sec:social}. Then, we compare the performance of FP, FPS, UPMR and UPMRS under social welfare maximization and peak pricing. The results are shown in Figure \ref{fig:peakwelfareenergyall}.
%
%Figure \ref{fig:peakwelfare} shows the normalized social welfare loss and energy bill with respect to the maximum ones among the four cases. Meanwhile, Figure \ref{fig:peakwelfareenergy1} and \ref{fig:peakwelfareenergy2} shows the detail of normalized energy usage of FP, FPS, UPMR and UPMRS with respect to the possible maximum one. Obviously, UPMR can effectively reduce the social welfare loss and energy bill without utilizing the storage system under electric rate of peak pricing. Meanwhile, we find that UPMRS can slightly reduce the welfare loss but increase the energy bill when comparing with UPMR. Besides, by comparing Figure \ref{fig:peakprofitall} and \ref{fig:peakwelfareenergyall}, we find that as a social planner, CDC can reduce energy bill and smooth CDC's energy usage more effectively than the one who aims at profit maximization.

\subsection{Under Time-Dependent Pricing}
We now turn to the evaluation of \ouralg under TDP. Specifically,
we consider that the CDC is charged through TDP based on energy consumption, where the prices are represented by the prices of Ameren real time price from January 1, 2014 to January 31, 2014 \cite{amerenrealtimeprice} as shown in Figure \ref{fig:energyprice}.

Figure \ref{fig:tdpprofitall} compares the results for UP, UPS, UPMR and UPMRS under profit maximization and TDP. We see that, under TDP, the CDC can obtain less profit gain improvement than the one under electricity rates of peak pricing via demand response. Specifically, CDC has less incentive to perform demand response under TDP. Utilizing the
energy storage usually cannot greatly improve the CDC's profit or reduce its energy bill. The reason is that, in most cases, the wear-and-tear cost of battery is higher than the extra profit that can be gained from charging/discharging the energy storage under TDP. Also, under TDP, the CDC can still effectively increase CDC's profit while reducing its energy bill via \ouralg.

\section{Conclusion}
We proposed
a simple yet effective pricing policy, called \ouralg, which,
 on top of the commonly-applied usage-based pricing policy,
 includes
 a reward component offered to tenants to allow a CDC operator to defer their workloads.
% With UPMR, users are provided with monetary incentives to grant CDC time-shift their workloads and CDC can modulate users' workloads based on SLA transparent to users to reduce its overall energy consumption/bill. We model user and CDC's control knobs separately. Specifically, the formulation of maximization of CDC's profit and social welfare are not convex. Thus,
For maximizing the CDC operator's profit through \ouralg, we presented an effective and low-complexity decomposition-based algorithm to optimize the reward
 rate.
With real-world workload traces, we evaluated \ouralg
in terms of maximizing the CDC's profit and reducing its energy cost,
against UP (a commonly-used pricing policy in today's CDCs).
 Our results showed that UPMR can obtain $4.9\%$ more profit gain and $12.9\%$ more energy bill reduction without utilizing the energy storage system.
 %Meanwhile, as a social planner, CDC can reduce its energy bill more effectively when comparing with one who aims at profit maximization.

\appendices
%\subsection{Proof of Theorem \ref{the:convex}}
%The optimization (\ref{pro:profit}) contains variable $\rho$, $m_{on}[t]$, $m_{of}[t]$, $S[t]$, $\phi_i[t]$ and $\eta_i[t]$. Apparently, the objective function defined by (\ref{equ:dcobjective})-(\ref{equ:hatlambda}) is concave and the inequality constraints (\ref{con:rholowerbound})-(\ref{con:phigeqeta}) are affine. Further, after initializing $\lfloor D_i \rfloor$ in (\ref{con:hatgeqnohat}) by a specified constant, the constraint (\ref{con:hatgeqnohat}) becomes affine, too. In this case, the optimization problem (\ref{pro:profit}) becomes convex \cite{Boyd2004}.
\section{Proof of Theorem \ref{the:unique}}\label{proo:unique}
Considering that $\lfloor D_i \rfloor$ is nondecreasing over $\rho$, Theorem \ref{the:unique} holds if and only if
\begin{equation}\label{equ:lbound=ubound}
\lfloor D_i \rfloor \left|_{\rho=\text{Ub}-\epsilon}\right.=D_i^{Lb}, \quad \forall i.
\end{equation}
We separately prove (\ref{equ:lbound=ubound}) when $\text{Lb} \geq \max_i\{(D_{max}+1)\kappa_i\}$ and $\text{Lb} < \max_i\{(D_{max}+1)\kappa_i\}$. Firstly, if $\text{Lb} \geq \max_i\{(D_{max}+1)\kappa_i\}$, then from (\ref{equ:dub}), we have
\begin{equation}\label{equ:ubinf}
\text{Ub}=\infty.
\end{equation}
Next, from (\ref{equ:dlb}), we get
\begin{equation}\label{equ:dilbequal}
D_i^{Lb}= D_{max}, \quad \forall i.
\end{equation}
From (\ref{equ:di}) and (\ref{equ:ubinf}), we get
\begin{equation}\label{equ:difloorl}
\lfloor D_i \rfloor \left|_{\rho=\text{Ub}-\epsilon}\right.=D_{max}, \quad \forall i.
\end{equation}
Thus, from (\ref{equ:dilbequal}) and (\ref{equ:difloorl}), (\ref{equ:lbound=ubound}) holds when $\text{Lb} \geq \max_i\{(D_{max}+1)\kappa_i\}$.

\noindent
When $\text{Lb}< \max_i\{(D_{max}+1)\kappa_i\}$, we have
\begin{equation}\label{equ:uboundstep5}
%\begin{aligned}
\text{Ub}=\min_{i \in \vartheta} \{(D_i^{Lb}+2)\kappa_i\}.
%\end{aligned}
\end{equation}
In this case, we prove the correctness of (\ref{equ:lbound=ubound}) by four steps:

\noindent
\textbf{Step 1}: For all $i \in \vartheta$, from (\ref{equ:uboundstep5}), we have
\begin{equation}\label{equ:uboundstep}
%\begin{aligned}
\text{Ub} \leq (D_i^{Lb}+2)\kappa_i, \quad \forall i \in \vartheta.
%\end{aligned}
\end{equation}
Bring (\ref{equ:uboundstep}) into (\ref{equ:di}), we have
\begin{equation}\label{equ:uboundstep1}
\lfloor D_i \rfloor \left|_{\rho=\text{Ub}-\epsilon}\right.\leq \left\lfloor \max\left(\min\left(D_i^{Lb}+1-\epsilon,D_{max}\right),0\right) \right\rfloor.
\end{equation}
From (\ref{equ:dlb}), we ensure that $D_i^{Lb}\geq 0$. In this case, $\min\left(D_i^{Lb}+1-\epsilon,D_{max}\right)\geq 0$. Combine it with (\ref{equ:uboundstep1}), we get
\begin{equation}\label{equ:uboundstep2}
\begin{aligned}
\lfloor D_i \rfloor \left|_{\rho=\text{Ub}-\epsilon}\right. &\leq \left\lfloor \min\left(D_i^{Lb}+1-\epsilon,D_{max}\right) \right\rfloor\\
& \leq \left\lfloor D_i^{Lb}+1-\epsilon \right\rfloor.\\
\end{aligned}
\end{equation}
Next, from (\ref{equ:dlb}), we ensure that $D_i^{Lb}$ is integer. In this case, $\left\lfloor D_i^{Lb}+1-\epsilon \right\rfloor=D_i^{Lb}$. Combine it with (\ref{equ:uboundstep2}), we ensure
\begin{equation}\label{con:upperbound1}
\begin{aligned}
\lfloor D_i \rfloor \left|_{\rho=\text{Ub}-\epsilon}\right. &\leq D_i^{Lb},  \quad \forall i \in \vartheta.\\
\end{aligned}
\end{equation}

\noindent
\textbf{Step 2}: For all $i \notin \vartheta$, from (\ref{equ:dlb}) and (\ref{equ:eee}), we have $D_i^{Lb}=D_{max}$, $\forall i \notin \vartheta$. From (\ref{equ:di}), we have $\lfloor D_i \rfloor \left|_{\rho=\text{Ub}-\epsilon}\right. \leq D_{max}$, $\forall i$. Therefore,
\begin{equation}\label{con:upperbound2}
\begin{aligned}
\lfloor D_i \rfloor \left|_{\rho=\text{Ub}-\epsilon}\right. &\leq D_i^{Lb},  \quad \forall i \notin \vartheta.\\
\end{aligned}
\end{equation}
Combine (\ref{con:upperbound1}) and (\ref{con:upperbound2}), we ensure
\begin{equation}\label{con:upperbound}
\begin{aligned}
\lfloor D_i \rfloor \left|_{\rho=\text{Ub}-\epsilon}\right. &\leq D_i^{Lb},  \quad \forall i.\\
\end{aligned}
\end{equation}

\noindent
\textbf{Step 3}: Let
\begin{equation}\label{ioptimal}
i^{'}=\arg \min_{i \in \vartheta} \{(D_i^{Lb}+2)\kappa_i\}.
\end{equation}
From (\ref{equ:uboundstep5}), we have
\begin{equation}\label{equ:ubounds}
\begin{aligned}
\text{Ub}&=(D_{i^{'}}^{Lb}+2)\kappa_{i^{'}}.\\
\end{aligned}
\end{equation}
Since $i^{'} \in \vartheta$, from (\ref{equ:eee}), we have Lb $\in [0,(D_{max}+1)\kappa_{i^{'}})$. %We divide the domain of Lb into two parts:
%
%\noindent
%1),
If Lb $\in [0,2\kappa_{i^{'}})$, from (\ref{equ:dlb}), $D_{i^{'}}^{Lb}=0$. Combine it with (\ref{equ:ubounds}), we have
\begin{equation}\label{equ:ubounds1}
\begin{aligned}
\text{Ub}&=2\kappa_{i^{'}}>\text{Lb}.
\end{aligned}
\end{equation}
%
%\noindent
If Lb $\in[2\kappa_{i^{'}},(D_{max}+1)\kappa_{i^{'}})$, from (\ref{equ:dlb}), $D_{i^{'}}^{Lb}=\lfloor \frac{\text{Lb}}{\kappa_{i^{'}}}-1\rfloor$. Combine it with (\ref{equ:ubounds}), we have
\begin{equation}\label{equ:ubounds2}
\begin{aligned}
\text{Ub}&=\left(\left\lfloor \frac{\text{Lb}}{\kappa_{i^{'}}}-1\right\rfloor+2\right)\kappa_{i^{'}}\\
&>\left(\left( \frac{\text{Lb}}{\kappa_{i^{'}}}-1\right)+1\right)\kappa_{i^{'}}\\
&>\text{Lb}.
\end{aligned}
\end{equation}
Combining (\ref{equ:ubounds1}) and (\ref{equ:ubounds2}), we ensure that Ub $>$ Lb. Meanwhile, since $\lfloor D_i \rfloor$ is nondecreasing over $\rho$, we have
\begin{equation}\label{con:lowerbound}
\begin{aligned}
\lfloor D_i \rfloor \left|_{\rho=\text{Ub}-\epsilon}\right. &\geq D_i^{Lb}, \quad \forall i.\\
\end{aligned}
\end{equation}

\noindent
\textbf{Step 4}: From (\ref{con:upperbound}) and (\ref{con:lowerbound}), (\ref{equ:lbound=ubound}) always holds when Lb $< \max_i\{(D_{max}+1)\kappa_i\}$, too.

\section{Proof of Theorem \ref{the:rho}}\label{proo:rho}
We notice that, the only part of the objective function of problem (\ref{equ:subproblem1}) that involves $\rho$ is the definition of the reward rate (\ref{equ:discountoverrho}). Meanwhile, (\ref{equ:discountoverrho}) involves no variable other than $\rho$. Also, the only constraint of problem (\ref{equ:subproblem1}) that involves $\rho$ is (\ref{con:newrholowerbound}), which involves no other constraint except $\rho$. Therefore, the problem (\ref{equ:subproblem1}) can be decomposed in two separate optimization problems, among which the first one is to minimize the Reward term defined by (\ref{equ:reward}) and (\ref{equ:discountoverrho}), subject to the constraints (\ref{equ:dlb})-(\ref{con:newrholowerbound}). The optimal value of $\rho$, in this decomposed optimization problem is $\rho=\text{Lb}$, since the Reward term defined by (\ref{equ:reward}) and (\ref{equ:discountoverrho}) is a non-decreasing function of $\rho$.

\end{document}